\documentclass[10pt,twocolumn]{article}

\usepackage{graphicx}
\usepackage{fancyhdr}
\RequirePackage{amsthm,amsmath,amsfonts,amssymb}
\RequirePackage[colorlinks,linkcolor=blue,citecolor=black,urlcolor=black]{hyperref}
\RequirePackage{graphicx}
\usepackage{natbib}
\RequirePackage[ruled,vlined]{algorithm2e}
\usepackage{enumerate}
\usepackage[T1]{fontenc}
\usepackage{booktabs}

\usepackage[margin=.65in]{geometry}

\RequirePackage[ruled,vlined]{algorithm2e}
\usepackage{enumerate}

\newlength\FHoffset
\fancyheadoffset{\FHoffset}
\title{\Large \bf Spiral-Elliptical automated galaxy morphology classification\\ from telescope images}
\author{Matthew J. Baumstark and Giuseppe Vinci\\{\small \it Department of Applied and Computational Mathematics and Statistics}\\ {\small \it University of Notre Dame, Crowley Hall, Notre Dame, Indiana, USA, E-mail: {\tt gvinci@nd.edu}}}
\date{}

\begin{document}

\maketitle

\small

\begin{abstract}
The classification of galaxy morphologies is an important step in the investigation of theories of hierarchical structure formation. While human expert visual classification remains quite effective and accurate, it cannot keep up with the massive influx of data from emerging sky surveys. A variety of approaches have been proposed to classify large numbers of galaxies; these approaches include crowdsourced visual classification, and automated and computational methods, such as machine learning methods based on designed morphology statistics and deep learning. In this work, we develop two novel galaxy morphology statistics, descent average and descent variance, which can be efficiently extracted from telescope galaxy images. We further propose simplified versions of the existing image statistics concentration, asymmetry, and clumpiness, which have been widely used in the literature of galaxy morphologies. We utilize the galaxy image data from the Sloan Digital Sky Survey to demonstrate the effective performance of our proposed image statistics at accurately detecting spiral and elliptical galaxies when used as features of a random forest classifier.
\end{abstract}

~


\section{Introduction}
\label{introduction}

Galaxy morphology identification is an important step in the study of galaxy structures and in testing theories of hierarchical structure formation in the universe \citep{conselice2003relationship,conselice2014evolution, blanton2009physical, shapley2011physical, van2011galaxy, silk2012current, conroy2013modeling}.  
Galaxies take on a variety of morphologies, but two shapes are of particular importance in dominant galaxy classification schemes \citep{hubble1937realm, de1959classification, sandage1961hubble,conselice2006fundamental,kormendy2011revised,graham2019galaxy}: spiral and elliptical. 

Spiral galaxies, like our own Milky Way, are galaxies that feature prominent star-filled arms radiating outward from their bright centers. Elliptical galaxies are galaxies in which stars are clustered around a bright center in an ellipsoidal geometry, appearing to observers as circular (or elliptical). Spiral galaxies are typically less massive and bluer with ongoing star formation, while elliptical galaxies are massive and red with limited star formation \citep{holmberg1958photographic}. Indeed, the study of spiral and elliptical shapes plays an important role in investigating the relationship between star formation and galaxy structure \citep{conselice2006fundamental, bauer2011star,kennicutt1998star,kennicutt2012star,madau2014cosmic}.

The structural differences between spiral and elliptical galaxies can generally be distinguished by eye with moderate effort. However, major advances in astrophotography have increased the number of observable galaxies by several orders of magnitude \citep{york2000sloan, gardner2006james, lintott2008galaxy,grogin2011candels}.  As a consequence, there are simply too many galaxies to classify manually. 
In the last two decades, researchers have attempted to identify galaxy morphologies using crowdsourced visual classification \citep{lintott2008galaxy}, or automated and computational methods, including machine learning methods using designed morphological statistics \citep{conselice2003relationship, lotz2004new, freeman2013new}, and deep learning \citep{vega2021pushing,walmsley2022galaxy,gupta2022galaxy,farias2020mask,barchi2020machine,dominguez2018improving,dieleman2015rotation,banerji2010galaxy,reza2021galaxy}. Deep learning classifiers can provide good classification accuracy, but these models are often difficult to interpret, and their implementation can be very computationally expensive. In this research, we revisit the more traditional approach to quantifying galaxy morphologies based on designed morphological statistics, which can be interpretable and computationally efficient, while still yielding excellent classification prediction accuracy. 

We develop two new interpretable galaxy morphology statistics, \textit{Descent Average} and \textit{Descent Variance}. Descent Average summarizes the distribution of light intensity across pixel rings radiating outward from the center of a galaxy, while Descent Variance summarizes the variability of light intensity within pixel rings. We further propose simplified versions of the three existing and widely used image statistics \textit{Concentration}, \textit{Asymmetry}, and \textit{Clumpiness} \citep{conselice2003relationship}. All five morphology statistics can be computed efficiently from the pixel intensities of telescope galaxy images. We utilize galaxy images from the Sloan Digital Sky Survey \citep{bowen1973optical, york2000sloan, gunn20062, smee2013multi, wilson2019apache, kollmeier2019sdss, perruchot2018integration} and demonstrate the effective performance of our proposed image statistics at accurately detecting spiral and elliptical galaxy shapes  when used as features of a random forest classifier. The classifier achieved a classification  accuracy of 95.5\% on test data.

This article is organized as follows. In Section~\ref{sec:stats}, we define our proposed image statistics, descent average, descent variance, and the simplified concentration, asymmetry, and clumpiness. In Section~\ref{sec:data}, we describe the training and the testing of our spiral-elliptical random forest classifier based on our proposed morphological statistics computed from galaxy images of the SDSS. Finally, in Section~\ref{sec:disc} we discuss our results and further describe our plans for future work.

\section{Galaxy shape statistics}\label{sec:stats}
In this section we propose five galaxy shape image statistics. Two of these five statistics, \textit{Descent Average} and \textit{Descent Variance}, are novel shape metrics, while the other three statistics are simplified versions of the \textit{Concentration}, \textit{Asymmetry}, and \textit{Clumpiness} (CAS) system \citep{conselice2003relationship}. This section is organized as follows. In Section~\ref{sec:rings} we describe \textit{ring segmentation}, a fundamental preprocessing step that is used to compute our image statistics. In Section~\ref{sec:dav}, we define the Descent Average and Descent Variance statistics. Finally, in Section~\ref{sec:cas} we present our simplified CAS statistics. Appendix~\ref{app:comp} provides details on the computational cost of these statistics.

\subsection{Ring segmentation}\label{sec:rings}
\begin{figure}
    \centering
    \includegraphics[width=\columnwidth]{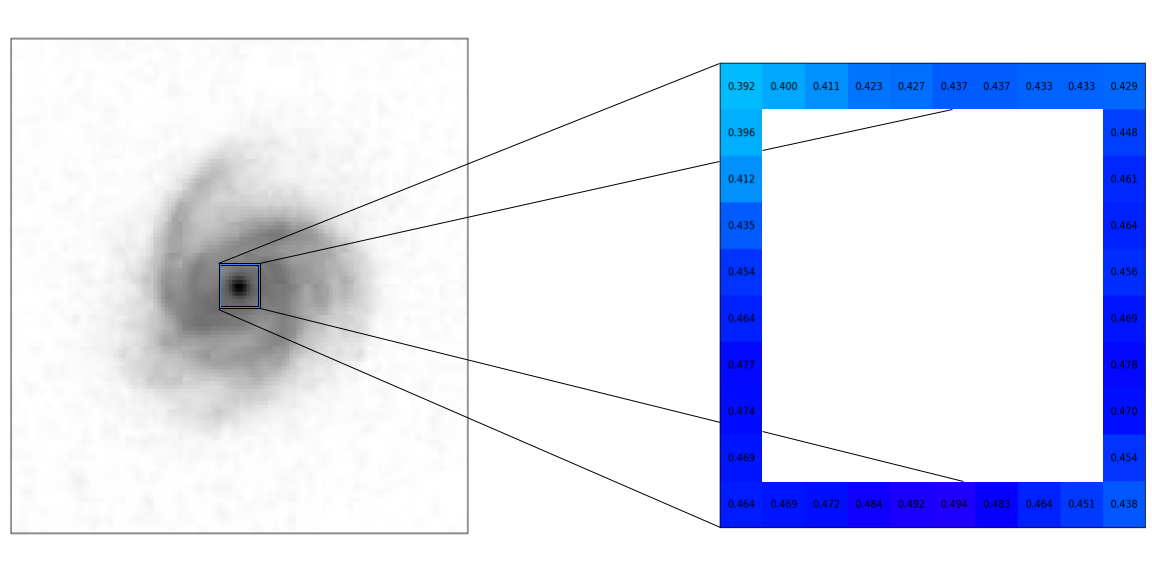}
    \caption{\small
        {\bf Ring segmentation}. A galaxy image is decomposed into concentric square rings of pixels. The plot highlights the pixel values contained in the tenth ring of the image of a spiral galaxy.
    }
    \label{fig:rings}
\end{figure}

Ring segmentation is an operation that decomposes a galaxy image into concentric groups of pixels. Specifically, a galaxy image is divided into a set of \textit{concentric square rings}, radiating outwards from the center of the galaxy. Given a $d\times d$ image matrix $I\in\mathbb{R}^{d\times d}_+$ of nonnegative light intensities, for $k=1,\ldots,d/2$, we define the $k$-th square ring of $I$ as 
\begin{equation}\label{eq:ring}
I^{(k)} = \{I_{ij}: (i,j)\in A_{d/2-k}\setminus A_{d/2-k+1}\}
\end{equation}
where $A_s = \{(i,j):1+s\le i,j\le d-s\}$ is the set of coordinates of the $(d-2s-1)\times (d-2s-1)$ central submatrix of $I$.  For example, if $d$ is even, then the first ring $I^{(1)}$ consists of the $2\times 2$ portion located at the center of the image. Then, the second ring $I^{(2)}$ is given by the set of pixels adjacently surrounding the first ring. Higher order rings can be sequentially identified analogously. For the case where $d$ is odd, the process is similar but with the first ring consisting of one central pixel. Figure \ref{fig:rings} illustrates the identification of the tenth square ring of a spiral galaxy image.

The concentric square rings defined above can be viewed as computationally efficient approximations of circular rings around the center of the galaxy, which carry information on the geometry of a galaxy shape. In Section~\ref{sec:data} we show that our approach based on square rings is nonetheless effective for galaxy shape classification. 

\subsection{Descent Average and Descent Variance}\label{sec:dav}
\begin{figure*}
    \centering
    \includegraphics[width=.85\textwidth]{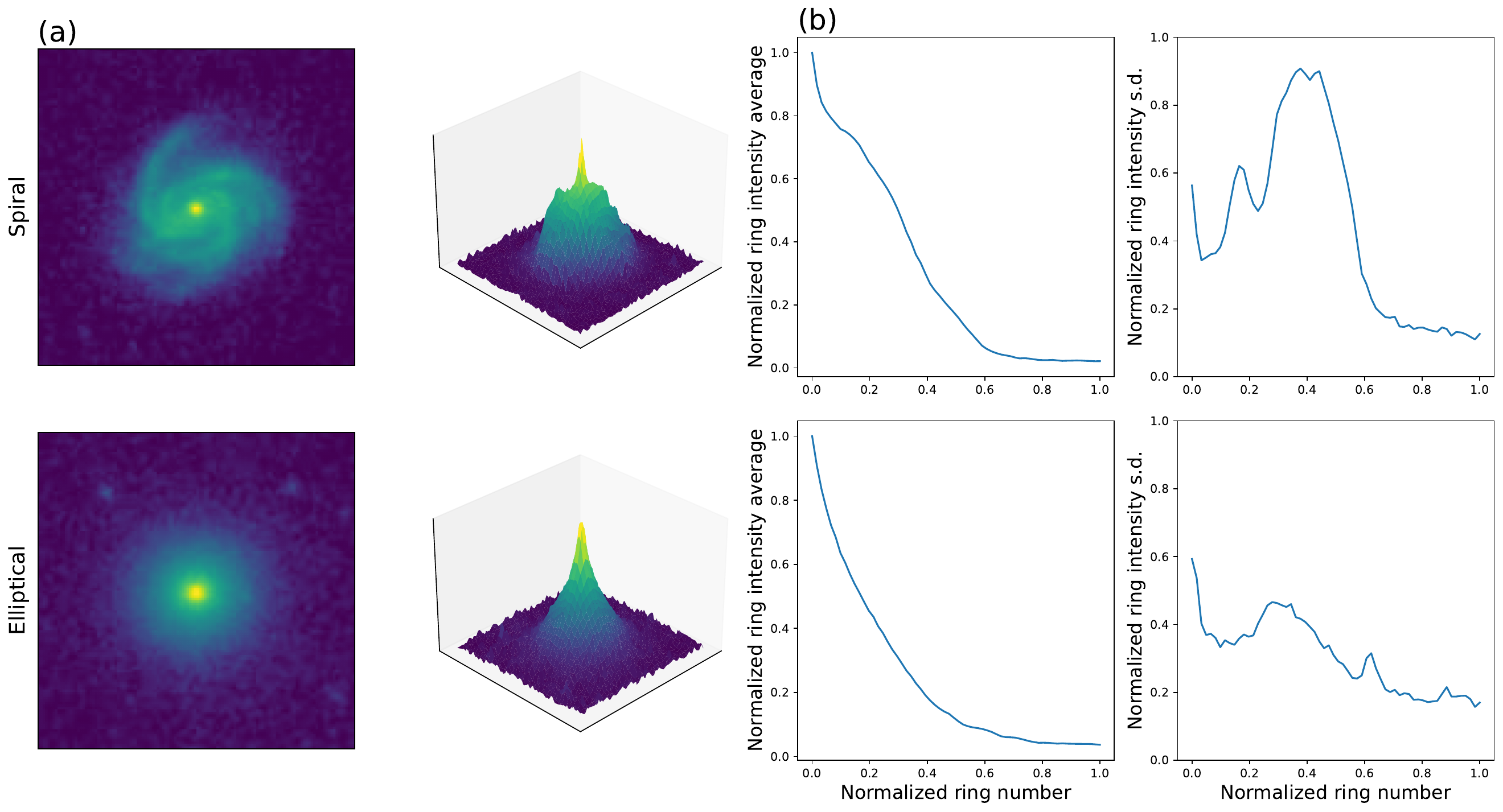}
    \caption{\small
        {\bf Computation of descent average and descent variance}. (a): Images of a spiral galaxy and an elliptical galaxy, and their corresponding surface maps. The spiral galaxy appears to retain more light in outer rings than the elliptical galaxy. (b): Normalized empirical mean and standard deviation of light intensity pixel values contained in the concentric square rings obtained by applying the ring segmentation process (Section~\ref{sec:rings}) on each galaxy image. The shape features descent average (DA) and descent variance (DV) are obtained as the integrals of the plotted curves. In this figure, the spiral galaxy has $DA=0.301$ and $DV=0.433$, while the elliptical galaxy has $DA=0.243$ and $DV=0.307$.
    }\vspace{-1mm}
    \label{fig:dav}
\end{figure*}

Figure~\ref{fig:dav}(a) shows the images of two spiral and elliptical galaxies, and their corresponding surface map representations, where the pixel intensity is plotted as a function of the pixel's coordinates. The figure shows an important difference in how the intensities descend from their centers. The surface map of the spiral galaxy has thick ``tails'', that is, the intensity profile holds substantial light in the outer regions of the galaxy. In contrast, the surface map of the elliptical galaxy features broadly smooth intensity descents from the galaxy center. Descent average and descent variance aim at measuring these properties of the light intensity distribution.

\subsubsection{Descent Average}\label{sec:da}
Descent average ($DA$) summarizes the distribution of light intensity \textit{across} rings radiating outward from the center of a galaxy. Specifically, $DA$ tries to captures the decay of average light intensity across different rings of different radii. The study of the average light intensity in galaxy rings has been given significant importance in parametric settings \citep{de1948recherches, de1959classification, sersic1963influence}, where the average light intensity profiles of elliptical galaxies are assumed to follow a specific parametric form whose decay is governed by a limited number of parameters; here, we take an empirical and nonparametric approach to summarize the light profile of a galaxy. 

To compute the $DA$ of a galaxy image $I\in\mathbb{R}_+^{d\times d}$, we first segment $I$ into $K = \lceil d/2 \rceil $ concentric square rings $I^{(1)},\ldots,I^{(K)}$, as described in Section~\ref{sec:rings}. Next, we compute the mean of the light intensity pixel values contained in each square ring: 
\begin{equation}
    \mu_k = {\rm mean}(I^{(k)})
\end{equation}
where, for any $w\in\mathbb{R}^n$, ${\rm mean}(w) = \frac{1}{n}\sum_{j=1}^n w_j$. Next, to eliminate any differences in overall magnitude across galaxies, the average intensities are normalized to range between 0 and 1, by dividing all values by their maximum. Finally, the $DA$ statistic is computed as the integral (or area) under the normalized ring intensity average curve plotted versus normalized ring number:
\begin{equation}
    DA = \frac{1}{K - 2} \sum_{i = 3}^K \frac{\mu_i}{\max_{j\ge 3}\mu_j},
\end{equation}
where the first two rings are ignored to mitigate the possible distorting effects of the higher light intensity on the bulge of the galaxy; eliminating the center is a common practice in galaxy feature computation (see \cite{conselice2003relationship}, as an example). Figure \ref{fig:dav}(b) shows the graphs of normalized ring intensity average as a function of normalized ring number for the spiral and elliptical galaxies shown in panel (a). The spiral galaxy image yields a curve with a slower decay and thereby a larger $DA$ compared with the elliptical galaxy.

\subsubsection{Descent Variance}\label{sec:dv}
Descent Variance ($DV$) summarizes the distribution of light intensity \textit{within} rings. Within a square ring, elliptical galaxies should not exhibit much variation in their pixel intensities. This is because of their largely smooth intensity descent curves (Figure \ref{fig:dav}(a)). However, spiral galaxies should have more variation - the arms of the galaxy extend into certain areas of the ring and do not in other areas, leading to higher variation. Therefore, computing the variance in each ring is another effective way of differentiating between spiral galaxies and elliptical galaxies. Using the same ring segmentation process outlined in Section~\ref{sec:rings}, the image $I\in\mathbb{R}_+^{d\times d}$ is divided into $K = \lceil d/2 \rceil $ concentric square rings $I^{(1)},\ldots,I^{(K)}$. The empirical standard deviation of the pixel intensities in each ring is then computed for all rings:
\begin{equation}
    {\rm sd}_k = {\rm sd}(I^{(k)}) 
\end{equation} 
where, for any $w\in\mathbb{R}^n$, ${\rm sd}(w) = \big(\frac{1}{n}\sum_{j=1}^n w_j^2-\big(\frac{1}{n}\sum_{j=1}^n w_j\big)^2\big)^{\frac{1}{2}}$. Next, we normalize ${\rm sd}_k$ by dividing it by $\alpha={\rm mean}(\tilde I)$, where $\tilde I$ is the pixel set excluding the first two rings. 
Finally, the $DV$ statistic is computed as the integral (or area) under the normalized standard deviation curve plotted versus normalized ring number:
\begin{equation}
    DV = \frac{1}{K - 2} \sum_{i = 3}^K \frac{{\rm sd}_i}{\alpha}
\end{equation}
where the first two rings are ignored to mitigate the possible distorting effects of the higher light intensity on the bulge of the galaxy, as explained in Section~\ref{sec:da}. Figure~\ref{fig:dav}(b) shows the graphs of normalized ring intensity standard deviation as a function of normalized ring number for the spiral and elliptical galaxies shown in panel (a). As mentioned earlier, because spiral arms lead to higher variances within the ring segments, the spiral galaxy image yields a standard deviation curve with larger values, and thereby a larger $DV$, compared with the elliptical galaxy.

\begin{figure}
    \centering
    \includegraphics[width=\columnwidth]{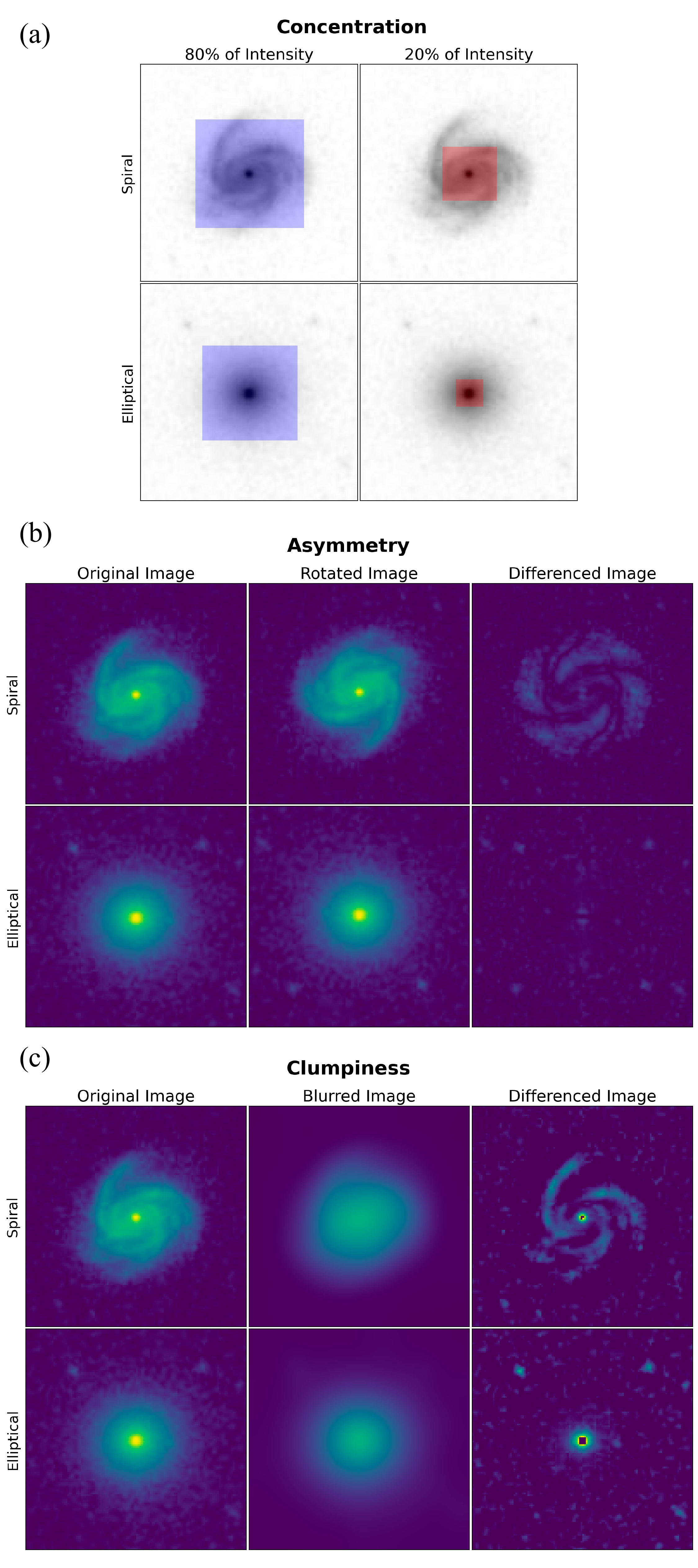}
    \caption{\small
        {\bf Computation of simplified concentration, asymmetry, and clumpiness.}     
        (a): Concentration ($C$) is the ratio of the size of the central square enclosing 80\% of the pixel intensity and that of the square enclosing 20\%. 
        (b): Asymmetry ($A$) is computed by summing the absolute entrywise differences of an image and its 180 degree rotated version.  
        (c): Clumpiness ($S$) is computed by summing the excess brightness of an image compared with a smoothed version of it. 
    } 
    \label{fig:cas}
\end{figure}

\subsection{Simplified CAS}\label{sec:cas}

The Concentration, Asymmetry, and Clumpiness (CAS) system \citep{conselice2003relationship} has proven to be a robust set of image statistics for galaxy morphology quantification. In this section we seek to implement simplified versions of CAS to use together with the two novel statistics $DA$ and $DV$ proposed in Section~\ref{sec:dav}. These streamlined versions of the CAS shape statistics are derived from the same methodologies and concepts used in the original CAS, but are simpler in their construction. It will be shown in Section~\ref{sec:data} that using these streamlined CAS statistics yield an accurate galaxy shape classification.

\subsubsection{Concentration}\label{sec:cas_conc}
The Concentration statistic of \cite{conselice2003relationship} is designed to measure how concentrated a galaxy's light profile is. The original statistic captures this by taking a ratio of the 80\% and 20\% growth radii. The streamlined version of concentration that we propose is based on a similar concept of taking a ratio of two radii enclosing different percentiles of the intensity in the image. To compute the radii, the ring segmentation process described in Section~\ref{sec:rings} is used to decompose the image $I\in\mathbb{R}_+^{d\times d}$ into $K=\lceil d/2\rceil$ concentric square rings $I^{(k)}$. Then, we define $i_{80}$ as the radius containing 80\% of the galaxy's intensity, that is, the index of the innermost ring $I^{(i_{80})}$ enclosing at least 80\% of the total intensity of the image, and the radius $i_{20}$ is defined analogously:
\begin{eqnarray}
    i_{80} &=& \min\{i:S_i\ge 0.8\}\\
    i_{20} &=& \min\{i:S_i\ge 0.2\}
\end{eqnarray}
where 
\begin{eqnarray}
    S_i &=& \frac{\sum_{k=1}^i s_k}{\sum_{j=1}^K s_j}\\
    s_k &=& \sum_j I^{(k)}_j
\end{eqnarray}
Finally, we define the simplified concentration as the ratio
\begin{equation}\label{eq:concentration}
    C=\frac{i_{80}}{i_{20}}
\end{equation}
\cite{conselice2003relationship} notes that elliptical galaxies should have higher  concentrations than spirals because more of their light is concentrated in the center of the galaxy. Figure~\ref{fig:cas}(a) depicts the computation of the simplified concentration statistics for a spiral galaxy and an elliptical galaxy.

\subsubsection{Asymmetry}\label{sec:cas_asy}
The Asymmetry statistic of \cite{conselice2003relationship} is designed to measure how asymmetrical a galaxy is. Following the conceptual idea underlying the original statistic, but without steps such as background noise removal, the image of the galaxy $I\in\mathbb{R}_+^{d\times d}$ is first rotated 180 degrees to produce $I^{180}$ with entries defined by
\begin{equation}
    I^{180}_{i,j} = I_{d-i+1,d-j+1}
\end{equation}
The rotated image is then subtracted from the original image element-wise and the absolute value is taken. This produces a differential image $D \in\mathbb{R}^{d\times d}$ that contains the asymmetrical features of the image:
\begin{equation}
    D_{i,j} = |I_{i,j} - I^{180}_{i,j}|
\end{equation}
Finally, the entries in $D$ are summed and divided by the sum of the pixel intensities in the original image. Therefore, the simplifed asymmetry statistic is defined by:
\begin{equation} \label{asymmetry}
    A = \frac{\sum_{i,j} D_{i,j}}{\sum_{i,j} I_{i,j}}
\end{equation}
Figure~\ref{fig:cas}(b) depicts the computation of the simplified asymmetry statistics for a spiral galaxy and an elliptical galaxy.

\subsubsection{Clumpiness}\label{sec:cas_clump} 
The Clumpiness statistic of \cite{conselice2003relationship} is designed to detect nonsmooth structures, which are typical of galaxies that are undergoing star formation, e.g.~spiral galaxies. Following the original statistic's underlying concept, but without steps such as background noise removal, the simplified version of clumpiness that we propose first employs the use of a Gaussian smoothing filter to eliminate high level image details. Smoothing filters are controlled by a parameter $\sigma$ that determines the level of blurring. \cite{conselice2003relationship} notes that any value of $\sigma$ should be able to detect the high-level features in a galaxy. We pick $\sigma$ to be sufficiently large ($\sigma = 7$). Let $I^{\sigma}$ denote the smoothed version of $I$. The blurred image is then subtracted from the original image, producing a differential image $D \in\mathbb{R}^{d\times d}$ that leaves the high-level structures in the galaxy:
\begin{equation}\label{eq:Diff}
    D_{i,j} = I_{i,j} - I^{\sigma}_{i,j}
\end{equation}
Following the recommendations of \cite{conselice2003relationship} here as well, all negative entries in $D$ and the ones corresponding to the center of the galaxy are forced to 0 (the center is approximated by identifying the top 1\% of pixel intensities in the differenced image). Finally, after these adjustments are completed, the remaining intensities in the differenced image are summed and are normalized by dividing by the sum of the intensities in the original image, yielding the simplified clumpiness statistic
\begin{equation} \label{eq:clumpiness}
    S = \frac{\sum_{i,j} D_{i,j}\cdot \mathcal{I}(0<D_{i,j}<q^*)}{\sum_{i,j} I_{i,j}}
\end{equation}
where $\mathcal{I}()$ is the indicator function and $q^*$ the 99\% empirical quantile of all intensity differences $\{D_{i,j}\}$ (Equation~\eqref{eq:Diff}). Figure~\ref{fig:cas}(c) depicts the computation of the simplified clumpiness statistics for a spiral galaxy and an elliptical galaxy.

\begin{figure*}
    \centering
    \includegraphics[width=1\textwidth]{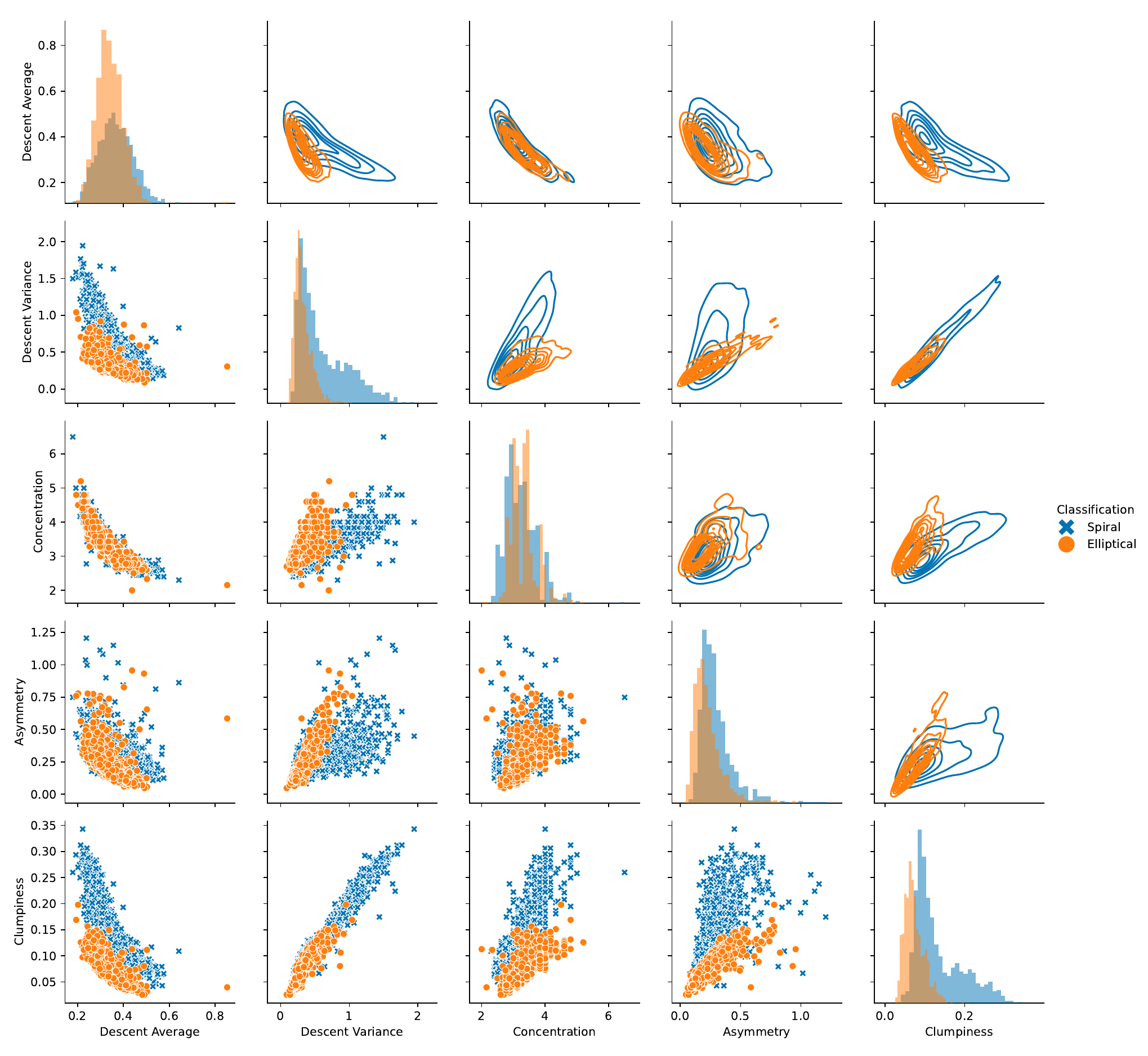}
    \caption{\small {\bf Exploratory data analysis}.  Histograms for the five shape statistics $DA,DV,C,A,S$ computed from the images of 1,000 spiral galaxies and 1,000 elliptical galaxies are displayed over the diagonal panels. Scatterplots of the statistics against one another are displayed in the lower triangular panels, while the contours of 2--dimensional kernel density estimates are shown on the upper triangular panels. Spiral galaxies appear to have higher levels of descent average, descent variance, asymmetry, and clumpiness, and lower levels of concentration. 
    }
    \label{fig:eda}
\end{figure*}

~

\section{Data analysis}\label{sec:data}
We now apply the shape statistics of Section~\ref{sec:stats} to the classification of galaxy images from the Sloan Digital Sky Survey (SDSS) \citep{bowen1973optical, york2000sloan, gunn20062, smee2013multi, wilson2019apache, kollmeier2019sdss, perruchot2018integration}. We focus on spiral and elliptical galaxies because of their prevalence in the known universe. However, we expect our statistics to be effective at identifying other galaxy shapes, such as lenticular and disturbed morphologies, and leave this exploration for future work. 
In Section~\ref{sec:datasumm} we describe the galaxy image data set and how we split it into training and testing set. In Section~\ref{sec:eda} we perform an exploratory analysis on a relatively small random subset of the data. In Section~\ref{sec:train} we train a random forest classifier based on our shape statistics, and then test it on the test data in Section~\ref{sec:test}.

\subsection{The SDSS data}\label{sec:datasumm}  The data set we consider contains images of 5,000 spiral galaxies and 5,000 elliptical galaxies (redshift $z\le 0.35$ for the vast majority), and with classification labels provided by the Galaxy Zoo \citep{lintott2008galaxy}. All images have resolution $64\times 64$ and were retrieved from the SDSS database with the procedure described in Appendix~\ref{app:sdss}.  We randomly split the data into a training set (80\%) and a test set (20\%). The training set is used to explore the data (Section~\ref{sec:eda}) and train our galaxy shape classifier (Section~\ref{sec:train}), while the test set is only used to test the performance of our finally trained classifier (Section~\ref{sec:test}).

\subsection{Exploratory data analysis} \label{sec:eda}
Before working with any supervised learning classification methods, we perform an exploratory data analysis on a subset of the training set to observe how well the proposed statistics (Section~\ref{sec:stats})  may perform at distinguishing spiral galaxies from elliptical galaxies. We randomly subsample 1,000 spiral galaxies and 1,000 elliptical galaxies from the training set and compute our five shape statistics for all 2,000 images. 

On the diagonal panels of Figure~\ref{fig:eda}, we show the histograms for the five shape statistics $DA,DV,C,A,S$ computed from the images of 1,000 spiral galaxies and 1,000 elliptical galaxies. Moreover, scatterplots of the statistics against one another are displayed in the lower triangular panels, while the contours of 2--dimensional kernel density estimates are shown on the upper triangular panels. We can see that, compared with elliptical galaxies, spiral galaxies appear to have higher levels of descent average, descent variance, asymmetry, and clumpiness, and lower levels of concentration. Among the five statistics, descent variance and the simplified clumpiness, individually and in particular jointly with descent average, appear to yield the strongest separation between distributions for spiral and elliptical galaxies. However, the five statistics can provide superior discriminative power when used jointly as features of a machine learning classifier, as shown in the following sections.

\subsection{Training the random forest classifier} \label{sec:train}
Several machine learning classification methods have been proposed and popularized in the last twenty years. Here, we employ a random forest model (\cite{breiman2001random}; R package randomForest) to perform galaxy morphology classification based on the shape statistics defined in Section~\ref{sec:stats}.

The random forest is a machine learning model built off of the decision tree, a simpler learning algorithm. Decision trees iterate through a set of data to locate the optimal split in the data; this is usually a split in a specific variable (for example,  ``if $\text{descent variance} \ge t$, then the galaxy is a spiral'', for some threshold $t$). This split is optimal in the sense that it leads to the largest information gain (or decrease of Gini impurity). Each split in the data creates a decision node, and this process is repeated after each split until the split leads to a perfect segmentation of the data, or until a prespecified maximum number of splits is achieved. Decision trees are easy to interpret: data points can be classified by following the decision tree down the ``branches'' until a classification is made. However, individual decision trees are unreliable because they are prone to overfitting, where the model essentially memorizes the data it is trained on, but cannot generalize its predictive power to unseen data.

Random forests mitigate the risk of overfitting by training a large number of decision trees and then combining them together to produce a unique decision scheme by majority rule. To ensure that the trees are not identical, random forests take advantage of bootstrap sampling and feature limitation. Each tree is trained on a bootstrap sample of the data (a sample of equal size taken with replacement), allowing each tree to form decisions based on different but identically distributed datasets. 
Additionally, each tree node split is implemented by employing only $m$ randomly selected features. This feature limitation further reduces the dependence between trees and thereby reduces variance of the ensemble model.

Random forests depend on various parameters, including the number of trees $T$, and the number of features $m$ at node splits. We set $T=500$, which is a typical default choice. The usual default choice for the number of features to use at node split is $m = \lfloor\sqrt{p}\rfloor$ \citep{hastie2009elements}, where $p$ is the total number of available features. However, we decide to select $m$ via 10-fold cross-validation (Appendix~\ref{app:cv}) to yield the best prediction accuracy. Furthermore, we consider four possible feature subsets: $(C,A,S)$, $(DA, DV)$, $(DA, DV,S)$, and $(DA, DV, C, A, S)$. In Table \ref{tab:cv} we summarize the results of cross-validation in the training data for the four feature sets and different values of $m=1,\ldots,5$. 
The setting that yielded the lowest cross-validated misclassification risk (proportion of misclassified cases averaged across ten folds) employs all five features ($DA, DV, C, A, S$) with $m=2$.  

\begin{table*}
    \centering
    \begin{tabular}[t]{lrrrr}
        \toprule
        & $(C,A,S)$ & $(DA,DV)$ & $(DA,DV,S)$ & $(DA,DV,C,A,S)$\\
        \midrule
        $m=1$ & 50.3 (6.3\%) & 94.3 (11.8\%) & 48.2 (6.0\%) & 36.2 (4.5\%) \\
        $m=2$ & 53.3 (6.7\%) & 97.8 (12.2\%) & 47.4 (5.9\%) & 35.5 (4.4\%) \\
        $m=3$ & 55.3 (6.9\%) & - & 48.1 (6.0\%) & 36.4 (4.6\%) \\
        $m=4$ & - & - & - & 36.2 (4.5\%)\\
        $m=5$ & - & - & - & 37.5 (4.7\%)\\
        \bottomrule
    \end{tabular}
    \caption{\small{\bf Random forest classifier prediction risk estimated from training data via 10-fold cross-validation}. The table shows the average number of misclassifications across 10-folds  and relative accuracy in parentheses for different feature sets and number of features $m$ to use in each tree split of the random forest. The setting with lowest risk involves all five features $(DA,DV,C,A,S)$ and use $m = 2$.
    }
    \label{tab:cv}
\end{table*}

\begin{figure}
    \centering
    \includegraphics[width=1.1\columnwidth]{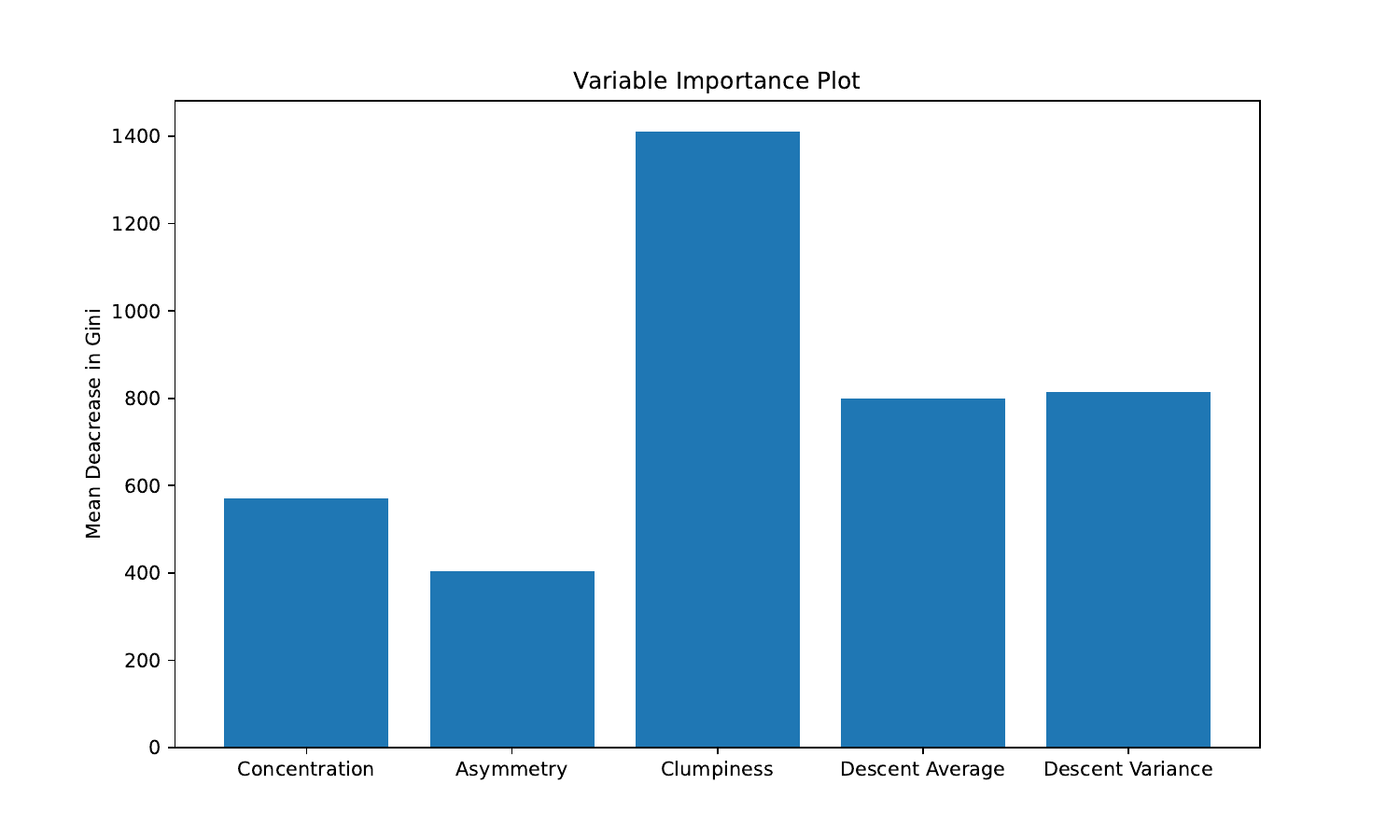}
    \caption{\small {\bf Variable importance of galaxy shape features in the final trained random forest classifier}. Clumpiness, descent variance and descent average are the most important variables, in agreement with the results of Table~\ref{tab:cv} where we can see that the combination $(DA,DV,S)$ produced a prediction risk close to the optimal one which involved all variables $(DA,DV,C,A,S)$. 
    }
    \label{fig:varimp}
\end{figure}

Finally, we use the full training data (8,000 galaxy images) to train our final classifier with the optimal setting identified via 10-fold cross-validation, that is using all statistics $(DA, DV, C, A, S)$ and $m=2$. Figure~\ref{fig:varimp} displays the variable importance of the five features. Variable importance quantifies how well each variable performs at increasing information gain (or decrease of Gini impurity) when used to determine a decision node. We can see that $DA$, $DV$, and $S$ have the largest importance in the random forest classifier, in agreement with our exploratory data analysis (Section~\ref{sec:eda}).

\subsection{Testing the random forest classifier}\label{sec:test}
We now test the performance of our final trained random forest classifier at predicting the shapes, spiral or elliptical, of the 2,000 galaxy images in the testing set. 

\begin{table}
    \centering
    \begin{tabular}[t]{lrr}
        \toprule
        & Actual Elliptical & Actual Spiral \\
        \midrule
        Predicted Elliptical & 984 & 54\\
        Predicted Spiral & 36 & 926\\
        \bottomrule
    \end{tabular}
    \caption{\small {\bf Confusion matrix of test set prediction}. The final trained random forest classifier produces 90 misclassifications, yielding a 95.50\% test prediction accuracy.
    }
    \label{tab:confusion}
\end{table}

\begin{figure}[t!]
    \centering
\includegraphics[width=1\columnwidth]{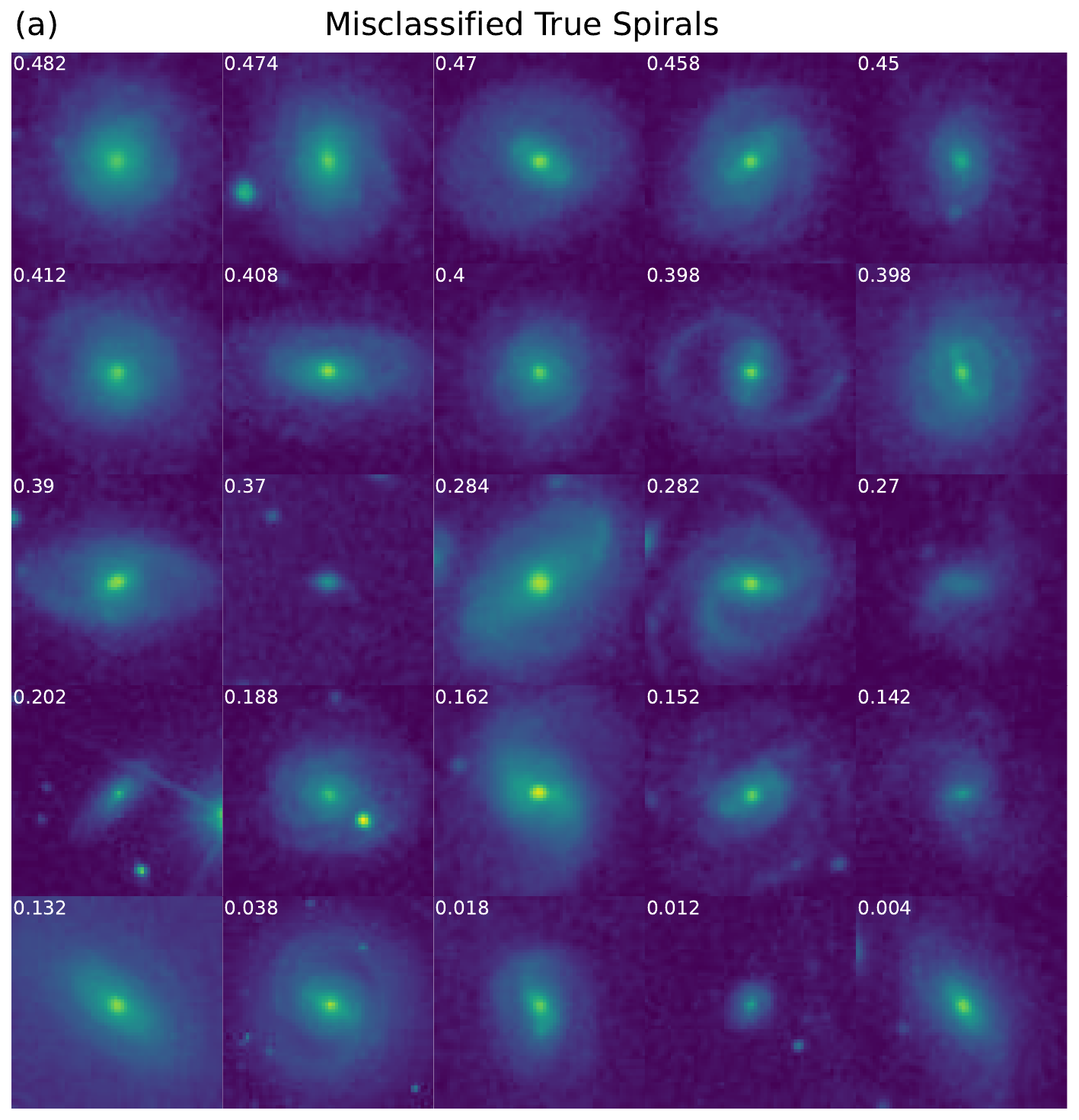}
     
\includegraphics[width=1\columnwidth]{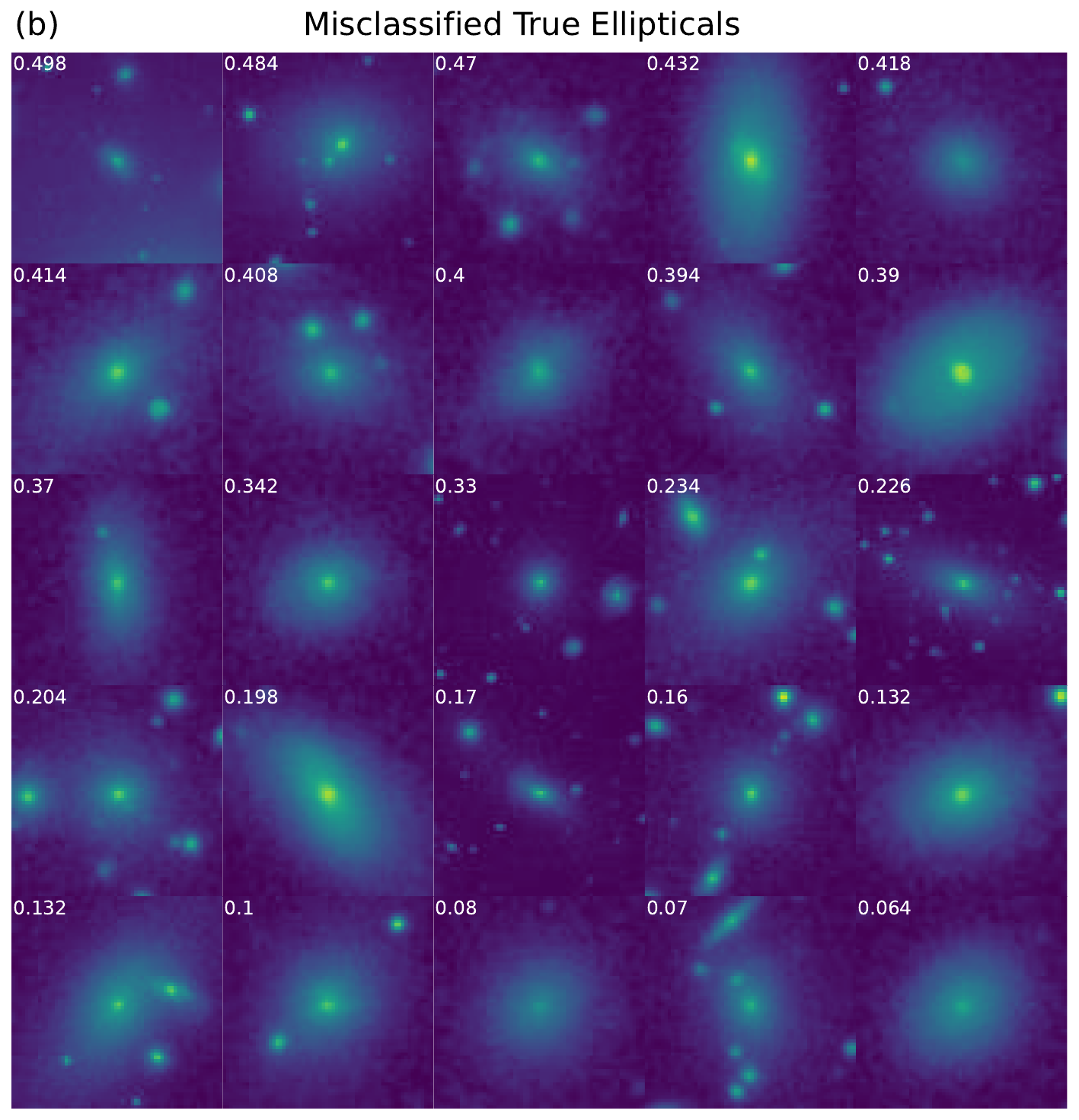}
   \caption{\small {\bf Examples of misclassified galaxy shapes in the test data.} (a): Subset of 25 misclassified spiral galaxies ordered by their probability of being spiral predicted by the random forest classifier. (b): Subset of 25 misclassified elliptical galaxies ordered by their probability of being elliptical predicted by the random forest classifier.} 
    \label{fig:misclass}
\end{figure}

Table \ref{tab:confusion} shows that the final random forest classifier makes 90 
misclassifications out of 2,000 galaxies in the testing set, implying a 95.50\% prediction accuracy rate. In Figure~\ref{fig:misclass} we visualize the images of 50 of the 90 misclassified galaxies. Many of the misclassified true spirals in panel (a) appear conform to similar circular geometries and many of the visually observable spiral features, such as arms, may be too faint for the statistics to detect. Background sources are an issue in both misclassification cases, but especially in the misclassified true ellipticals. 
Indeed, the presence of noisy peaks of light in the image can inflate the statistics to levels that would lead the model to predict the galaxy as a spiral. 
Altogether, most misclassifications in the testing set appear to be due to the presence of strong noise peaks of light and faint imagery.

\section{Discussion}\label{sec:disc}
We proposed two new galaxy morphology statistics, Descent Average and Descent Variance. Descent Average summarizes the distribution of light intensity across rings radiating outward from the center of a galaxy, while Descent Variance summarizes the variability of light intensity within rings. Spiral galaxies typically yield higher values of Descent Average and Descent Variance than elliptical galaxies. We further proposed simplified versions of the three existing and widely used image statistics Concentration, Asymmetry, and Clumpiness (Conselice, 2003). We used all five statistics to train a random forest classifier for automated classification of galaxies into spirals and ellipticals, the two predominant galaxy shapes in the known universe. Our classifier yielded a 95.50\% classification accuracy on test data.

For future work, we will continue to improve the accuracy of our statistics computation by refining the image preprocessing steps, such as denoising, galaxy centering, and rescaling, thereby improving the discriminant power of our classifier. Moreover, we will investigate the robustness of our shape statistics at challenging levels of redshift. This future study may lead us to enrich our statistics with appropriate bias corrections for redshift effects, which typically consist of morphological distortions and image degradation \citep{conselice2014evolution}. This investigation can be conducted through massive simulations \citep{giavalisco1996morphology, conselice2003relationship}, in which the morphological statistics computed at the original redshift levels of the images are compared with those computed on higher redshift perturbed simulated versions of the galaxy images, allowing us to identify appropriate bias terms to subtract from our statistics given the level of redshift.

Furthermore, we expect our statistics to be effective at identifying other galaxy shapes, such as the spiral sub-classes with or without bars \citep{de1959classification}, 
and disturbed morphologies, including galaxy mergers and irregular galaxies \citep{freeman2013new,conselice2006galaxy,jogee2009history,de2007millennium,ellison2010galaxy,lotz2004new}. Indeed, our shape statistics could be nested into a larger multiclass classification problem together with other existing shape statistics, such as the Gini and $M_{20}$ statistics \citep{lotz2004new}, and the Multimode, Intensity and Deviation (MID) statistics \citep{freeman2013new}, which were found especially useful to detect galaxy mergers. Our statistics may also be useful in the estimation of the distribution of galaxy morphologies via unsupervised learning 
\citep{schutter2015galaxy,vinci2014estimating}.

Finally, we are eager to apply our methodology in the analysis of massive simulated data  \citep{torrey2015synthetic,snyder2015galaxy,snyder2019automated,holincheck2016galaxy,rodriguez2017role,dickinson2018galaxy} and data from ongoing large scale sky surveys produced by the James Webb Space Telescope \citep{gardner2006james}.

\section*{Acknowledgements}
This research was funded by the Beutter Family Endowment for Excellence in Applied Mathematics and Computer Science, University of Notre Dame. 

Funding for the Sloan Digital Sky Survey V has been provided by the Alfred P. Sloan Foundation, the Heising-Simons Foundation, the National Science Foundation, and the Participating Institutions. SDSS acknowledges support and resources from the Center for High-Performance Computing at the University of Utah. The SDSS web site is \text{www.sdss.org}.

SDSS is managed by the Astrophysical Research Consortium for the Participating Institutions of the SDSS Collaboration, including the Carnegie Institution for Science, Chilean National Time Allocation Committee (CNTAC) ratified researchers, the Gotham Participation Group, Harvard University, Heidelberg University, The Johns Hopkins University, L’Ecole polytechnique f\'{e}d\'{e}rale de Lausanne (EPFL), Leibniz-Institut f\"{u}r Astrophysik Potsdam (AIP), Max-Planck-Institut f\"{u}r Astronomie (MPIA Heidelberg), Max-Planck-Institut f\"{u}r Extraterrestrische Physik (MPE), Nanjing University, National Astronomical Observatories of China (NAOC), New Mexico State University, The Ohio State University, Pennsylvania State University, Smithsonian Astrophysical Observatory, Space Telescope Science Institute (STScI), the Stellar Astrophysics Participation Group, Universidad Nacional Aut\'{o}noma de M\'{e}xico, University of Arizona, University of Colorado Boulder, University of Illinois at Urbana-Champaign, University of Toronto, University of Utah, University of Virginia, Yale University, and Yunnan University.

\appendix

\section{The SDSS data} \label{app:sdss}
The Sloan Digital Sky Survey (SDSS) \citep{bowen1973optical, york2000sloan, gunn20062, smee2013multi, wilson2019apache, kollmeier2019sdss, perruchot2018integration} is one of the most-cited and largest sky surveys ever completed, and its data are publicly available (www.sdss.org). The scope of the data is vast, ranging from spectra to images of all types of objects: stars, supernovae, and, importantly for this research, galaxies. 

To obtain galaxy images from the SDSS database, we interact with two different entities within the SDSS data access tool SkyServer (skyserver.sdss.org): the Catalog Archive Server Jobs System (CasJobs) and the Image Cutout (ImgCutout) service \citep{nieto2004imgcutout}. CasJobs allows users to employ Structured Query Language (SQL), a common database query language, to obtain customized lists of galaxies within the SDSS, along with their astronomical coordinates (right ascension and declination), a measure of their size in the night sky (e.g. $r^{90}$, the radius containing 90\% of the galaxy's Petrosian flux \citep{petrosian1976surface}), and classification labels from Galaxy Zoo \citep{lintott2008galaxy}. The ImgCutout service can then be used to retrieve the images of the galaxies in the list obtained via CasJobs, based on their astronomical coordinates, desired image dimensions ($d\times d$), and scale (arcseconds per pixel), which we compute as
\begin{equation}
    {\rm scale} = \frac{2 r^{90}}{dp}
\end{equation}
where $p$ is the pixel size. In our analyses we used $d=64$ and $p=5/6$, and required $r^{90}\ge 10$. The full procedure can be implemented in Python and R using the libraries available at \text{http://casjobs.sdss.org/casjobs/casjobscl.aspx}.

\section{Cross-validation}\label{app:cv}
Cross-validation is a procedure that allows us to approximate the test prediction error of supervised learning methods by using the training data. A common version of cross-validation is $K$--fold cross-validation \citep{hastie2009elements}, which is implemented as follows. We first evenly split the training set into $K$ disjoint subsets $T_1,\ldots,T_K$. Then, for each data subset $T_k$, the supervised learning model is trained on all of the data except $T_k$, and then the trained model is used to predict the labels in $T_k$ given the predictors in $T_k$. We finally compute the classification errors for $K$ models and average them to obtain a final approximation of the prediction error. In our analyses we use $K=10$, which is a common choice in machine learning.

\section{Computational time}\label{app:comp}
All computations were implemented in Python and R on a 2018 MacBook Pro laptop, with processor 2.3 GHz Quad-Core Intel Core i5, and memory 8 GB 2133 MHz LPDDR3. The computation of all five shape statistics $(DA, DV, C, A, S)$ for 10,000 images took only about 32.59 seconds, i.e. about 0.003259 seconds per image.


\bibliographystyle{elsarticle-harv} 
\bibliography{refs.bib}

\begin{thebibliography}{58}
\expandafter\ifx\csname natexlab\endcsname\relax\def\natexlab#1{#1}\fi
\expandafter\ifx\csname url\endcsname\relax
  \def\url#1{\texttt{#1}}\fi
\expandafter\ifx\csname urlprefix\endcsname\relax\def\urlprefix{URL }\fi

\bibitem[{Banerji et~al.(2010)Banerji, Lahav, Lintott, Abdalla, Schawinski,
  Bamford, Andreescu, Murray, Raddick, Slosar, et~al.}]{banerji2010galaxy}
Banerji, M., Lahav, O., Lintott, C.~J., Abdalla, F.~B., Schawinski, K.,
  Bamford, S.~P., Andreescu, D., Murray, P., Raddick, M.~J., Slosar, A.,
  et~al., 2010. Galaxy zoo: reproducing galaxy morphologies via machine
  learning. Monthly Notices of the Royal Astronomical Society 406~(1),
  342--353.

\bibitem[{Barchi et~al.(2020)Barchi, de~Carvalho, Rosa, Sautter, Soares-Santos,
  Marques, Clua, Gon{\c{c}}alves, de~S{\'a}-Freitas, and
  Moura}]{barchi2020machine}
Barchi, P.~H., de~Carvalho, R., Rosa, R.~R., Sautter, R., Soares-Santos, M.,
  Marques, B.~A., Clua, E., Gon{\c{c}}alves, T., de~S{\'a}-Freitas, C., Moura,
  T., 2020. Machine and deep learning applied to galaxy morphology-a
  comparative study. Astronomy and Computing 30, 100334.

\bibitem[{Bauer et~al.(2011)Bauer, Conselice, P{\'e}rez-Gonz{\'a}lez,
  Gr{\"u}tzbauch, Bluck, Buitrago, and Mortlock}]{bauer2011star}
Bauer, A.~E., Conselice, C.~J., P{\'e}rez-Gonz{\'a}lez, P.~G., Gr{\"u}tzbauch,
  R., Bluck, A.~F., Buitrago, F., Mortlock, A., 2011. Star formation in a
  stellar mass-selected sample of galaxies to z= 3 from the goods-nicmos
  survey. Monthly Notices of the Royal Astronomical Society 417~(1), 289--303.

\bibitem[{Blanton and Moustakas(2009)}]{blanton2009physical}
Blanton, M.~R., Moustakas, J., 2009. Physical properties and environments of
  nearby galaxies. Annual Review of Astronomy and Astrophysics 47, 159--210.

\bibitem[{Bowen and Vaughan(1973)}]{bowen1973optical}
Bowen, I., Vaughan, A., 1973. The optical design of the 40-in. telescope and of
  the irenee dupont telescope at las campanas observatory, chile. Applied
  Optics 12~(7), 1430--1435.

\bibitem[{Breiman(2001)}]{breiman2001random}
Breiman, L., 2001. Random forests. Machine learning 45~(1), 5--32.

\bibitem[{Conroy(2013)}]{conroy2013modeling}
Conroy, C., 2013. Modeling the panchromatic spectral energy distributions of
  galaxies. Annual Review of Astronomy and Astrophysics 51, 393--455.

\bibitem[{Conselice(2003)}]{conselice2003relationship}
Conselice, C.~J., 2003. The relationship between stellar light distributions of
  galaxies and their formation histories. The Astrophysical Journal Supplement
  Series 147~(1), 1.

\bibitem[{Conselice(2006{\natexlab{a}})}]{conselice2006fundamental}
Conselice, C.~J., 2006{\natexlab{a}}. The fundamental properties of galaxies
  and a new galaxy classification system. Monthly Notices of the Royal
  Astronomical Society 373~(4), 1389--1408.

\bibitem[{Conselice(2006{\natexlab{b}})}]{conselice2006galaxy}
Conselice, C.~J., 2006{\natexlab{b}}. Galaxy mergers and interactions at high
  redshift. Proceedings of the International Astronomical Union 2~(S235),
  381--384.

\bibitem[{Conselice(2014)}]{conselice2014evolution}
Conselice, C.~J., 2014. The evolution of galaxy structure over cosmic time.
  Annual Review of Astronomy and Astrophysics 52, 291--337.

\bibitem[{De~Propris et~al.(2007)De~Propris, Conselice, Liske, Driver, Patton,
  Graham, and Allen}]{de2007millennium}
De~Propris, R., Conselice, C.~J., Liske, J., Driver, S.~P., Patton, D.~R.,
  Graham, A.~W., Allen, P.~D., 2007. The millennium galaxy catalogue: The
  connection between close pairs and asymmetry; implications for the galaxy
  merger rate. The Astrophysical Journal 666~(1), 212.

\bibitem[{de~Vaucouleurs(1948)}]{de1948recherches}
de~Vaucouleurs, G., 1948. Recherches sur les nebuleuses extragalactiques. In:
  Annales d'Astrophysique. Vol.~11. p. 247.

\bibitem[{de~Vaucouleurs(1959)}]{de1959classification}
de~Vaucouleurs, G., 1959. Classification and morphology of external galaxies.
  In: Astrophysik iv: Sternsysteme/astrophysics iv: Stellar systems. Springer,
  pp. 275--310.

\bibitem[{Dickinson et~al.(2018)Dickinson, Fortson, Lintott, Scarlata, Willett,
  Bamford, Beck, Cardamone, Galloway, Simmons, et~al.}]{dickinson2018galaxy}
Dickinson, H., Fortson, L., Lintott, C., Scarlata, C., Willett, K., Bamford,
  S., Beck, M., Cardamone, C., Galloway, M., Simmons, B., et~al., 2018. Galaxy
  zoo: morphological classification of galaxy images from the illustris
  simulation. The Astrophysical Journal 853~(2), 194.

\bibitem[{Dieleman et~al.(2015)Dieleman, Willett, and
  Dambre}]{dieleman2015rotation}
Dieleman, S., Willett, K.~W., Dambre, J., 2015. Rotation-invariant
  convolutional neural networks for galaxy morphology prediction. Monthly
  notices of the royal astronomical society 450~(2), 1441--1459.

\bibitem[{Dom{\'\i}nguez~S{\'a}nchez et~al.(2018)Dom{\'\i}nguez~S{\'a}nchez,
  Huertas-Company, Bernardi, Tuccillo, and Fischer}]{dominguez2018improving}
Dom{\'\i}nguez~S{\'a}nchez, H., Huertas-Company, M., Bernardi, M., Tuccillo,
  D., Fischer, J., 2018. Improving galaxy morphologies for sdss with deep
  learning. Monthly Notices of the Royal Astronomical Society 476~(3),
  3661--3676.

\bibitem[{Ellison et~al.(2010)Ellison, Patton, Simard, McConnachie, Baldry, and
  Mendel}]{ellison2010galaxy}
Ellison, S.~L., Patton, D.~R., Simard, L., McConnachie, A.~W., Baldry, I.~K.,
  Mendel, J.~T., 2010. Galaxy pairs in the sloan digital sky survey--ii. the
  effect of environment on interactions. Monthly Notices of the Royal
  Astronomical Society 407~(3), 1514--1528.

\bibitem[{Farias et~al.(2020)Farias, Ortiz, Damke, Arancibia, and
  Solar}]{farias2020mask}
Farias, H., Ortiz, D., Damke, G., Arancibia, M.~J., Solar, M., 2020. Mask
  galaxy: Morphological segmentation of galaxies. Astronomy and Computing 33,
  100420.

\bibitem[{Freeman et~al.(2013)Freeman, Izbicki, Lee, Newman, Conselice,
  Koekemoer, Lotz, and Mozena}]{freeman2013new}
Freeman, P., Izbicki, R., Lee, A., Newman, J., Conselice, C., Koekemoer, A.,
  Lotz, J., Mozena, M., 2013. New image statistics for detecting disturbed
  galaxy morphologies at high redshift. Monthly Notices of the Royal
  Astronomical Society 434~(1), 282--295.

\bibitem[{Gardner et~al.(2006)Gardner, Mather, Clampin, Doyon, Greenhouse,
  Hammel, Hutchings, Jakobsen, Lilly, Long, et~al.}]{gardner2006james}
Gardner, J.~P., Mather, J.~C., Clampin, M., Doyon, R., Greenhouse, M.~A.,
  Hammel, H.~B., Hutchings, J.~B., Jakobsen, P., Lilly, S.~J., Long, K.~S.,
  et~al., 2006. The james webb space telescope. Space Science Reviews 123,
  485--606.

\bibitem[{Giavalisco et~al.(1996)Giavalisco, Livio, Bohlin, Macchetto, and
  Stecher}]{giavalisco1996morphology}
Giavalisco, M., Livio, M., Bohlin, R.~C., Macchetto, F.~D., Stecher, T.~P.,
  1996. On the morphology of the hst faint galaxies. The Astronomical Journal
  112, 369.

\bibitem[{Graham(2019)}]{graham2019galaxy}
Graham, A.~W., 2019. A galaxy classification grid that better recognizes
  early-type galaxy morphology. Monthly Notices of the Royal Astronomical
  Society 487~(4), 4995--5009.

\bibitem[{Grogin et~al.(2011)Grogin, Kocevski, Faber, Ferguson, Koekemoer,
  Riess, Acquaviva, Alexander, Almaini, Ashby, et~al.}]{grogin2011candels}
Grogin, N.~A., Kocevski, D.~D., Faber, S., Ferguson, H.~C., Koekemoer, A.~M.,
  Riess, A.~G., Acquaviva, V., Alexander, D.~M., Almaini, O., Ashby, M.~L.,
  et~al., 2011. Candels: the cosmic assembly near-infrared deep extragalactic
  legacy survey. The Astrophysical Journal Supplement Series 197~(2), 35.

\bibitem[{Gunn et~al.(2006)Gunn, Siegmund, Mannery, Owen, Hull, Leger, Carey,
  Knapp, York, Boroski, et~al.}]{gunn20062}
Gunn, J.~E., Siegmund, W.~A., Mannery, E.~J., Owen, R.~E., Hull, C.~L., Leger,
  R.~F., Carey, L.~N., Knapp, G.~R., York, D.~G., Boroski, W.~N., et~al., 2006.
  The 2.5 m telescope of the sloan digital sky survey. The Astronomical Journal
  131~(4), 2332.

\bibitem[{Gupta et~al.(2022)Gupta, Srijith, and Desai}]{gupta2022galaxy}
Gupta, R., Srijith, P., Desai, S., 2022. Galaxy morphology classification using
  neural ordinary differential equations. Astronomy and Computing 38, 100543.

\bibitem[{Hastie et~al.(2009)Hastie, Tibshirani, Friedman, and
  Friedman}]{hastie2009elements}
Hastie, T., Tibshirani, R., Friedman, J.~H., Friedman, J.~H., 2009. The
  elements of statistical learning: data mining, inference, and prediction.
  Vol.~2. Springer.

\bibitem[{Holincheck et~al.(2016)Holincheck, Wallin, Borne, Fortson, Lintott,
  Smith, Bamford, Keel, and Parrish}]{holincheck2016galaxy}
Holincheck, A.~J., Wallin, J.~F., Borne, K., Fortson, L., Lintott, C., Smith,
  A.~M., Bamford, S., Keel, W.~C., Parrish, M., 2016. Galaxy zoo:
  Mergers--dynamical models of interacting galaxies. Monthly Notices of the
  Royal Astronomical Society 459~(1), 720--745.

\bibitem[{Holmberg(1958)}]{holmberg1958photographic}
Holmberg, E., 1958. A photographic photometry of extragalactic nebulae.
  Meddelanden fran Lunds Astronomiska Observatorium Serie II 136, 1.

\bibitem[{Hubble(1937)}]{hubble1937realm}
Hubble, E., 1937. The realm of the nebulae. Ciel et Terre, Vol. 53, p. 194 53,
  194.

\bibitem[{Jogee et~al.(2009)Jogee, Miller, Penner, Skelton, Conselice,
  Somerville, Bell, Zheng, Rix, Robaina, et~al.}]{jogee2009history}
Jogee, S., Miller, S.~H., Penner, K., Skelton, R.~E., Conselice, C.~J.,
  Somerville, R.~S., Bell, E.~F., Zheng, X.~Z., Rix, H.-W., Robaina, A.~R.,
  et~al., 2009. History of galaxy interactions and their impact on star
  formation over the last 7 gyr from gems. The Astrophysical Journal 697~(2),
  1971.

\bibitem[{Kennicutt~Jr(1998)}]{kennicutt1998star}
Kennicutt~Jr, R.~C., 1998. Star formation in galaxies along the hubble
  sequence. Annual Review of Astronomy and Astrophysics 36~(1), 189--231.

\bibitem[{Kennicutt~Jr and Evans(2012)}]{kennicutt2012star}
Kennicutt~Jr, R.~C., Evans, N.~J., 2012. Star formation in the milky way and
  nearby galaxies. Annual Review of Astronomy and Astrophysics 50, 531--608.

\bibitem[{Kollmeier et~al.(2019)Kollmeier, Anderson, Blanc, Blanton, Covey,
  Crane, Drory, Frinchaboy, Froning, Johnson, et~al.}]{kollmeier2019sdss}
Kollmeier, J., Anderson, S., Blanc, G., Blanton, M., Covey, K., Crane, J.,
  Drory, N., Frinchaboy, P., Froning, C., Johnson, J., et~al., 2019. Sdss-v
  pioneering panoptic spectroscopy. Bulletin of the American Astronomical
  Society.

\bibitem[{Kormendy and Bender(2011)}]{kormendy2011revised}
Kormendy, J., Bender, R., 2011. A revised parallel-sequence morphological
  classification of galaxies: structure and formation of s0 and spheroidal
  galaxies. The Astrophysical Journal Supplement Series 198~(1), 2.

\bibitem[{Lintott et~al.(2008)Lintott, Schawinski, Slosar, Land, Bamford,
  Thomas, Raddick, Nichol, Szalay, Andreescu, et~al.}]{lintott2008galaxy}
Lintott, C.~J., Schawinski, K., Slosar, A., Land, K., Bamford, S., Thomas, D.,
  Raddick, M.~J., Nichol, R.~C., Szalay, A., Andreescu, D., et~al., 2008.
  Galaxy zoo: morphologies derived from visual inspection of galaxies from the
  sloan digital sky survey. Monthly Notices of the Royal Astronomical Society
  389~(3), 1179--1189.

\bibitem[{Lotz et~al.(2004)Lotz, Primack, and Madau}]{lotz2004new}
Lotz, J.~M., Primack, J., Madau, P., 2004. A new nonparametric approach to
  galaxy morphological classification. The Astronomical Journal 128~(1), 163.

\bibitem[{Madau and Dickinson(2014)}]{madau2014cosmic}
Madau, P., Dickinson, M., 2014. Cosmic star-formation history. Annual Review of
  Astronomy and Astrophysics 52, 415--486.

\bibitem[{Nieto-Santisteban et~al.(2004)Nieto-Santisteban, Szalay, and
  Gray}]{nieto2004imgcutout}
Nieto-Santisteban, M.~A., Szalay, A.~S., Gray, J., 2004. Imgcutout, an engine
  of instantaneous astronomical discovery. In: Astronomical Data Analysis
  Software and Systems (ADASS) XIII. Vol. 314. p. 666.

\bibitem[{Perruchot et~al.(2018)Perruchot, Guy, Le~Guillou, Blanc, Ronayette,
  R{\'e}gal, Castagnoli, Sepulveda, Le~van Suu, Jullo,
  et~al.}]{perruchot2018integration}
Perruchot, S., Guy, J., Le~Guillou, L., Blanc, P.-E., Ronayette, S., R{\'e}gal,
  X., Castagnoli, G., Sepulveda, E., Le~van Suu, A., Jullo, E., et~al., 2018.
  Integration and testing of the desi multi-object spectrograph: performance
  tests and results for the first unit out of ten. In: Ground-based and
  Airborne Instrumentation for Astronomy VII. Vol. 10702. SPIE, pp. 2294--2312.

\bibitem[{Petrosian(1976)}]{petrosian1976surface}
Petrosian, V., 1976. Surface brightness and evolution of galaxies. The
  Astrophysical Journal 209, L1--L5.

\bibitem[{Reza(2021)}]{reza2021galaxy}
Reza, M., 2021. Galaxy morphology classification using automated machine
  learning. Astronomy and Computing 37, 100492.

\bibitem[{Rodriguez-Gomez et~al.(2017)Rodriguez-Gomez, Sales, Genel, Pillepich,
  Zjupa, Nelson, Griffen, Torrey, Snyder, Vogelsberger,
  et~al.}]{rodriguez2017role}
Rodriguez-Gomez, V., Sales, L.~V., Genel, S., Pillepich, A., Zjupa, J., Nelson,
  D., Griffen, B., Torrey, P., Snyder, G.~F., Vogelsberger, M., et~al., 2017.
  The role of mergers and halo spin in shaping galaxy morphology. Monthly
  Notices of the Royal Astronomical Society 467~(3), 3083--3098.

\bibitem[{Sandage(1961)}]{sandage1961hubble}
Sandage, A., 1961. The Hubble atlas of galaxies. Vol. 618. Carnegie Institution
  of Washington Washington, DC.

\bibitem[{Schutter and Shamir(2015)}]{schutter2015galaxy}
Schutter, A., Shamir, L., 2015. Galaxy morphology---an unsupervised machine
  learning approach. Astronomy and Computing 12, 60--66.

\bibitem[{S{\'e}rsic(1963)}]{sersic1963influence}
S{\'e}rsic, J., 1963. Influence of the atmospheric and instrumental dispersion
  on the brightness distribution in a galaxy. Boletin de la Asociacion
  Argentina de Astronomia La Plata Argentina 6, 41--43.

\bibitem[{Shapley(2011)}]{shapley2011physical}
Shapley, A.~E., 2011. Physical properties of galaxies from z= 2--4. Annual
  Review of Astronomy and Astrophysics 49, 525--580.

\bibitem[{Silk and Mamon(2012)}]{silk2012current}
Silk, J., Mamon, G.~A., 2012. The current status of galaxy formation. Research
  in Astronomy and Astrophysics 12~(8), 917.

\bibitem[{Smee et~al.(2013)Smee, Gunn, Uomoto, Roe, Schlegel, Rockosi, Carr,
  Leger, Dawson, Olmstead, et~al.}]{smee2013multi}
Smee, S.~A., Gunn, J.~E., Uomoto, A., Roe, N., Schlegel, D., Rockosi, C.~M.,
  Carr, M.~A., Leger, F., Dawson, K.~S., Olmstead, M.~D., et~al., 2013. The
  multi-object, fiber-fed spectrographs for the sloan digital sky survey and
  the baryon oscillation spectroscopic survey. The Astronomical Journal
  146~(2), 32.

\bibitem[{Snyder et~al.(2019)Snyder, Rodriguez-Gomez, Lotz, Torrey, Quirk,
  Hernquist, Vogelsberger, and Freeman}]{snyder2019automated}
Snyder, G.~F., Rodriguez-Gomez, V., Lotz, J.~M., Torrey, P., Quirk, A.~C.,
  Hernquist, L., Vogelsberger, M., Freeman, P.~E., 2019. Automated distant
  galaxy merger classifications from space telescope images using the illustris
  simulation. Monthly Notices of the Royal Astronomical Society 486~(3),
  3702--3720.

\bibitem[{Snyder et~al.(2015)Snyder, Torrey, Lotz, Genel, McBride,
  Vogelsberger, Pillepich, Nelson, Sales, Sijacki, et~al.}]{snyder2015galaxy}
Snyder, G.~F., Torrey, P., Lotz, J.~M., Genel, S., McBride, C.~K.,
  Vogelsberger, M., Pillepich, A., Nelson, D., Sales, L.~V., Sijacki, D.,
  et~al., 2015. Galaxy morphology and star formation in the illustris
  simulation at z= 0. Monthly Notices of the Royal Astronomical Society
  454~(2), 1886--1908.

\bibitem[{Torrey et~al.(2015)Torrey, Snyder, Vogelsberger, Hayward, Genel,
  Sijacki, Springel, Hernquist, Nelson, Kriek, et~al.}]{torrey2015synthetic}
Torrey, P., Snyder, G.~F., Vogelsberger, M., Hayward, C.~C., Genel, S.,
  Sijacki, D., Springel, V., Hernquist, L., Nelson, D., Kriek, M., et~al.,
  2015. Synthetic galaxy images and spectra from the illustris simulation.
  Monthly Notices of the Royal Astronomical Society 447~(3), 2753--2771.

\bibitem[{Van~der Kruit and Freeman(2011)}]{van2011galaxy}
Van~der Kruit, P., Freeman, K., 2011. Galaxy disks. Annual Review of Astronomy
  and Astrophysics 49, 301--371.

\bibitem[{Vega-Ferrero et~al.(2021)Vega-Ferrero, Dom{\'\i}nguez~S{\'a}nchez,
  Bernardi, Huertas-Company, Morgan, Margalef, Aguena, Allam, Annis, Avila,
  et~al.}]{vega2021pushing}
Vega-Ferrero, J., Dom{\'\i}nguez~S{\'a}nchez, H., Bernardi, M.,
  Huertas-Company, M., Morgan, R., Margalef, B., Aguena, M., Allam, S., Annis,
  J., Avila, S., et~al., 2021. Pushing automated morphological classifications
  to their limits with the dark energy survey. Monthly Notices of the Royal
  Astronomical Society 506~(2), 1927--1943.

\bibitem[{Vinci et~al.(2014)Vinci, Freeman, Newman, Wasserman, and
  Genovese}]{vinci2014estimating}
Vinci, G., Freeman, P., Newman, J., Wasserman, L., Genovese, C., 2014.
  Estimating the distribution of galaxy morphologies on a continuous space.
  Proceedings of the International Astronomical Union 10~(S306), 68--71.

\bibitem[{Walmsley et~al.(2022)Walmsley, Lintott, G{\'e}ron, Kruk, Krawczyk,
  Willett, Bamford, Kelvin, Fortson, Gal, et~al.}]{walmsley2022galaxy}
Walmsley, M., Lintott, C., G{\'e}ron, T., Kruk, S., Krawczyk, C., Willett,
  K.~W., Bamford, S., Kelvin, L.~S., Fortson, L., Gal, Y., et~al., 2022. Galaxy
  zoo decals: Detailed visual morphology measurements from volunteers and deep
  learning for 314 000 galaxies. Monthly Notices of the Royal Astronomical
  Society 509~(3), 3966--3988.

\bibitem[{Wilson et~al.(2019)Wilson, Hearty, Skrutskie, Majewski, Holtzman,
  Eisenstein, Gunn, Blank, Henderson, Smee, et~al.}]{wilson2019apache}
Wilson, J., Hearty, F., Skrutskie, M., Majewski, S., Holtzman, J., Eisenstein,
  D., Gunn, J., Blank, B., Henderson, C., Smee, S., et~al., 2019. The apache
  point observatory galactic evolution experiment (apogee) spectrographs.
  Publications of the Astronomical Society of the Pacific 131~(999), 055001.

\bibitem[{York et~al.(2000)York, Adelman, Anderson~Jr, Anderson, Annis,
  Bahcall, Bakken, Barkhouser, Bastian, Berman, et~al.}]{york2000sloan}
York, D.~G., Adelman, J., Anderson~Jr, J.~E., Anderson, S.~F., Annis, J.,
  Bahcall, N.~A., Bakken, J., Barkhouser, R., Bastian, S., Berman, E., et~al.,
  2000. The sloan digital sky survey: Technical summary. The Astronomical
  Journal 120~(3), 1579.

\end{thebibliography}

\end{document}